\documentclass[usegraphicx,usenatbib]{mn2e}
\renewcommand\[{\begin{equation}}
\renewcommand\]{\end{equation}}
\def\parn{\par\noindent}

\def\arctan{{\rm arctan}}
\def\arctanh{{\rm arctanh}}
\def\As{A_1}

\def\Ass{A_{11}}
\def\Ash{A_{12}}

\def\eps{{\cal E}}
\def\epu{\epsilon_1}

\def\eulB{{\rm B}}
\def\fid{f_{\rm k}}
\def\ftil{\tilde f}
\def\Ftil{\tilde F}
\def\Ftilis{\Ftil_{\rm i}}
\def\Ftilan{\Ftil_{\rm a}}
\def\fn{f_{\rm N}}
\def\fis{f_{\rm i}}
\def\fan{f_{\rm a}}

\def\gau{\gamma_1}
\def\gad{\gamma_2}
\def\Is{I_1}

\def\Iss{I_{11}}
\def\Ish{I_{12}}

\def\ln{\hbox{${\rm ln}\, $}}

\def\parn{\par\noindent}

\def\psin{\Psi_{\rm N}}
\def\psit{\Psi_{\rm T}}

\def\psii{\Psi_{\rm k}}
\def\psitil{\tilde\Psi}

\def\qi{Q_{\rm k}}

\def\ms{M_1}
\def\mt{M_{\rm T}}

\def\ngam{\hbox{$n$-$\gamma$}}
\def\ngamma{\hbox{$n$-$\gamma$ }}
\def\ngaugad{\hbox{($n_1$-$\gau$,$n_2$-$\gad$) }}

\def\Qtil{\tilde Q}
\def\racp{r_{\rm a}^+}
\def\racm{r_{\rm a}^-}
\def\ra{r_{\rm a}}
\def\rai{r_{\rm ak}}
\def\rc{r_{\rm c}}
\def\rcu{r_{\rm c1}}
\def\rcd{r_{\rm c2}}
\def\rh{r_0}
\def\rs{r_1}

\def\rhos{\rho_1}
\def\rhon{\rho_{\rm N}}

\def\roi{\rho _{\rm k}}
\def\roqi{\varrho _{\rm k}}

\def\sa{s_{\rm a}}

\def\sacp{s_{\rm a}^+}
\def\sacm{s_{\rm a}^-}
\def\sM{s_{\rm M}}
\def\sigr{\sigma_{\rm r}}
\def\sigt{\sigma_{\rm t}}
\def\wsh{W_{12}}

\catcode`\@=11
\def\gsim{\ifmmode{\mathrel{\mathpalette\@versim>}}
    \else{$\mathrel{\mathpalette\@versim>}$}\fi}
\def\lsim{\ifmmode{\mathrel{\mathpalette\@versim<}}
    \else{$\mathrel{\mathpalette\@versim<}$}\fi}
\def\@versim#1#2{\lower 2.9truept \vbox{\baselineskip 0pt \lineskip
    0.5truept \ialign{$\m@th#1\hfil##\hfil$\crcr#2\crcr\sim\crcr}}}
\catcode`\@=12

   \title[Two--component galaxy models]
   {Two--component galaxy models: 
       the effect of density profile at large radii on the phase--space
       consistency}

   \author[Ciotti \& Morganti]
           {Luca Ciotti \& Lucia Morganti\thanks{Current address:
 Max-Planck-Institut f\"ur Ex. Physik, Giessenbachstra\ss{}e,
 D-85741 Garching, Germany}\\Astronomy Department, 
      University of Bologna, via Ranzani 1, 40127 Bologna, Italy}

\date{Submitted, October 1, 2008. 
Accepted version, November 3, 2008.}
\usepackage[fleqn]{amsmath}
\usepackage{amsfonts}

\begin{document} 
\maketitle

\begin{abstract} 
  It is well known that the density and anisotropy profile in the
  inner regions of a stellar system with positive phase-space
  distribution function are not fully independent.  Here we study the
  interplay between density profile and orbital anisotropy at large
  radii in physically admissible (consistent) stellar systems. The
  analysis is carried out by using two-component (\ngam,$\gau$)
  spherical self-consistent galaxy models, in which one density
  distribution follows a generalized $\gamma$ profile with external
  logarithmic slope $n$, and the other a standard $\gau$ profile (with
  external slope 4). The two density components have different
    ``core'' radii, the orbital anisotropy is controlled with the
    Osipkov-Merritt recipe, and for simplicity we assume that the mass
    of the $\gau$ component dominates the total potential everywhere.
  The necessary and sufficient conditions for phase-space consistency
  are determined analytically, also in presence of a dominant massive
  central black hole, and the analytical phase-space distribution
  function of (\ngam,1) models, and of \ngamma models with a central
  black hole, is derived for $\gamma=0,1,2$. It is found that the
  density slope in the external regions of a stellar system can play
  an important role in determining the amount of admissible
  anisotropy: in particular, for fixed density slopes in the central
  regions, systems with a steeper external density profile can support
  more radial anisotropy than externally flatter models. This is
    quantified by an inequality formally identical to the ``cusp
    slope-central anisotropy'' theorem (An \& Evans 2006), relating at
    all radii (and not just at the center) the density logarithmic
    slope and the anisotropy indicator in all Osipkov-Merritt
    systems.
\end{abstract}


\section{Introduction}

Observationally it is well established that elliptical galaxies
  have dark matter halos, and also host central supermassive black
  holes. These empirical facts motivate the study of multi-component
  dynamical models. When studying dynamical models of stellar systems
(single or multi-component), the minimal requirement to be met by a
physically acceptable model is the positivity of the phase-space
distribution function (DF) of each distinct component.  A model
satisfying this essential requirement (which is much weaker than
stability, but stronger than the fact that the Jeans equations have a
physically acceptable solution) is called {\it consistent}; moreover,
when the total gravitational potential is determined by the total
density profile through the Poisson equation, the model is called {\it
  self--consistent}. In other words, we call self--consistent a
consistent self-gravitating system.

Two general strategies can be used to construct a (self) consistent
model, or check whether a proposed model is (self) consistent: they
are commonly referred to as the ``$f$--to--$\rho$'' and the
``$\rho$--to--$f$'' approaches, where $f$ is the model DF (e.g., see
Bertin 2000, Binney \& Tremaine 2008).  An example of the first
approach is the survey of self--consistent two--component galaxy
models carried out by Bertin and co--workers, where the stellar and
dark matter components are described by two DFs of the $f_{\infty}$
family (e.g., Bertin \& Stiavelli 1984, Bertin et al. 1992); other
well known examples are the King (1966) models and the $f_{\nu}$
models (Bertin \& Trenti 2003).  Unfortunately, the resulting spatial
densities obtained by solving the associated Poisson equation are in
general not expressible in terms of simple or even known functions,
and so only numerical investigations are usually feasible.  In the
``$\rho$--to--$f$'' approach the density distribution is given, and
specific assumptions about the model internal dynamics are made; in
special cases inversion formulae from the density to the DF can
be obtained, usually in integral form or series expansion (see, e.g.,
Fricke 1952, Lynden--Bell 1962, Osipkov 1979, Merritt 1985, hereafter
OM; Dejonghe 1986, 1987; Cuddeford 1991; Hunter \& Qian 1993, Ciotti
\& Bertin 2005).  In particular, in order to recover the DF of
spherical models with orbital anisotropy, the OM technique has been
developed from the Eddington (1916) inversion for isotropic systems,
and widely used to study one and two--component models (see, e.g.,
Ciotti \& Pellegrini 1992, hereafter CP92; Hiotelis 1994; Carollo et
al. 1995; Ciotti \& Lanzoni 1997; Ciotti 1996, 1999, hereafter C96,
C99; Baes \& Dejonghe 2004; Buyle et al. 2007).  We remark that the OM
parameterization is not necessarily the best description of real
systems, however its simplicity and the fact that it captures the main
features of models of galaxy formation, that are generally found
nearly isotropic at the center and increasingly radially anisotropic
in the outer envelope (e.g., van Albada 1982; Trenti, Bertin \& van
Albada 2005; Nipoti, Londrillo \& Ciotti 2006; Binney \& Tremaine
2008. But see Cuesta et al. 2008 and references therein), make it the
natural choice for investigations as that presented in this paper.

In many cases, the difficulties inherent in the operation of
recovering analytically the DF prevent a simple consistency analysis,
and phase--space positivity must be investigated by numerical
inspection of the inversion integral. In these cases the reasons
underlying consistency or inconsistency of a proposed model tend to be
obscured by the numerical nature of the solution.  Fortunately,
informations about consistency of multi--component OM systems can be
obtained without recovering their DF, following the procedure
described in CP92.  This method uses the radial density profile of
each component and the total potential of the system, and gives
necessary and sufficient conditions for (self) consistency.  Moreover,
since only spatial differentiation and inequality checks are required,
this method is best suited for analytical investigations.  For
example, C96 and C99 applied the CP92 technique to the general family
of two--component, spherically symmetric and radially anisotropic
$(\gau,\gad)$ models.  This family is made of the superposition of two
$\gamma$ models (Dehnen 1993, Tremaine et al. 1994) with different
total masses, scale--lengths, inner density slopes, and OM radially
anisotropic velocity dispersions. The possibility to investigate the
combined effects of radial anisotropy and inner density slope on
multi--component systems made the study of $(\gau,\gad)$ models
interesting, as it is well known that the inner density profile sets
important constraints on the amount of admissible radial anisotropy
(e.g., Richstone \& Tremaine 1984), and indeed in C96 and C99
analytical limitations on anisotropy as a function of the density
slopes $\gau$ and $\gad$ were obtained.  These models clarified
the reasons behind the numerical findings of CP92, i.e. the difficulty
of consistently superimposing a centrally peaked distribution such as
the de Vaucouleurs (1948) profile to a centrally flat one, such as the
King (1972) or quasi--isothermal density profile (even in the
isotropic case).  In fact, it was shown that the DF of the $\gau$
component in isotropic $(\gau,\gad)$ models is nowhere negative,
independently of the mass and concentration of the $\gad$ component,
whenever $0\leq\gad\leq\gau$ and $1\leq\gau <3$.  On the contrary, a
$\gau=0$ component becomes inconsistent when adding a $\gad=1$ halo
with a small core radius. Thus, in two--component isotropic models,
the component with the steeper central density distribution is usually
the most robust against inconsistency.  More recently, the importance of
the central density slope in limiting the amount of possible radial
anisotropy has also been quantified with the so called ``cusp
slope-central anisotropy'' theorem (An \& Evans 2006, hereafter AE06;
see also eq.~[28] in de Bruijne et al. 1996).

The previous investigations left however unexplored the importance of
the \textit{external} density slope in determining the model
consistency.  In fact, the phase--space density cannot be identified,
in general, with any specific spatial position in the system, as (for
example) stars of a given energy can span a large range of radial
positions\footnote{Incidentally, this implies that the use of
  $\rho/\sigma^3$ as a proxy for phase--space density, where $\sigma$
  is the local value of the velocity dispersion, has no assignable
  meaning without an appropriate discussion.  For example, in
  power--law isotropic systems with $\rho\propto r^{-\gamma}$, because
  the functions $\rho/\sigma^3\propto r^{\frac{\gamma-6}{2}}$ (for
  $\gamma>1$) and phase--space density
  $f(\eps)\propto\eps^{\frac{6-\gamma}{2(\gamma-2)}}$ (for $\gamma>2$,
  where $\eps=-E$ is the so--called relative energy, see Sect.~4 and
  Baes et al. 2005) are both power laws with respect to their
  arguments, the exponents are related in a simple way. However, the
  converse statement is not true: for example, in the Plummer (1911)
  sphere $f(\eps)\propto\eps^{7/2}$ is a power law at all energies,
  but $\rho/\sigma^3$ is not a power law of radius.}  (systems made of
circular orbits are an obvious exception).  Therefore also the
external regions of a density distribution can be important in
limiting the maximum allowable anisotropy, but the $(\gau,\gad)$
models are of no help in the study of this issue, because the external
density profiles of both components all decrease as $r^{-4}$.  For
these reasons here we focus on the phase--space properties of \ngamma
models, i.e. models similar to the standard $\gamma$ models in the
inner regions, but with a density profile proportional to $r^{-n}$
(instead of $r^{-4}$) in the external regions; remarkably, several
properties of \ngamma models can be obtained from those of $\gamma$
models by differentiation with respect to their scale--length.  In
this notation, the 4-$\gamma$ models coincide with the standard
$\gamma$ models.  We also study the larger class of ($\ngam$,$\gau$)
models, i.e. two--component systems in which a $\gau$ halo is added to
a \ngamma component.  Thus, here we further explore the trends emerged
in CP92, C96, and C99, determining the limits imposed by phase--space
consistency on the parameters describing (\ngam,$\gau$) models, and
\ngamma models with a central BH [hereafter (\ngam,BH) models], with
particular focus on the effects of the external slope parameter
$n$. In specific cases (that we call {\it halo-dominated}
  models), the calculations are performed under the assumption that
  the mass of the halo component (or of the central BH) is dominant
  over the mass of the visible one.  This assumption is mainly
  motivated by mathematical simplicity (see also Sect.~4), although
  this is not the only reason.  In fact, for any given two--component
  model, it is expected that the DF properties are bracketed by those
  of the one component model and by those of the halo-dominated model
  (corresponding to the formal case of an infinite halo mass). Of
  course, while the case of dark matter dominated systems can be
  considered a viable representation of some real astrophysical
  systems, the case of a dominant BH is less natural, and it just
  gives the strongest possible limitations for consistency of
  systems with a central BH.

The paper is organized as follows.  In Section 2 we briefly review the
technique developed in CP92, and we prove that the necessary condition
for consistency derived in CP92 for OM systems can be rewritten {\it
  exactly} as the AE06 ``cusp slope-central anisotropy'' theorem,
holding however at all radii and not just at the center.  In Section 3
the one and two--component \ngamma models are introduced, and the
necessary and sufficient conditions imposed on the model parameters by
phase--space consistency are derived for different values of the
logarithmic density slope $n$. In Section 4 the DF of the \ngamma
component of halo-dominated ($\ngam$,1) models with
$\gamma=0,1,2$, and of similar models with a dominant central
BH, are derived explicitly for arbitrary (but integer) values of $n$,
and the true boundaries of the consistency region in the parameter
space are obtained. The main results of the investigation are
summarized in Section 5.  Finally, in the Appendix an easy method to
solve analytically the Jeans equations in the general OM case for the
wider class of ($n_1$-$\gau$,$n_2$-$\gad$) models is presented.

\section{The consistency of multi--component OM systems}

A stellar system made of the sum of the density components
$\roi$ is called consistent if each DF ($\fid$) is non--negative over
the whole accessible phase--space; a consistent self--gravitating system
is called self--consistent.  The technique developed in CP92 permits
us to check whether the DF of a multi--component spherical system,
where the orbital anisotropy of each component is modeled according to
the OM parameterization, is positive; in practice, only information
about the radial trend of each density component and of the total
integrated mass are used. In the OM formulation the DF of each
component is obtained assuming $\fid=\fid(\qi)$, with
\begin{equation}\label{eq:Q}
\qi=\eps-\frac{J^2}{2\rai^2},
\end{equation}
where $\eps$ and $J$ are respectively the relative energy and the
angular momentum modulus per unit mass, $\rai$ is the so--called {\it
  anisotropy radius}, and $\fid=0$ for $\qi\leq 0$ (e.g. see Binney \&
Tremaine 2008).  The velocity dispersion tensor associated with
eq.~(\ref{eq:Q}) is characterized by radial anisotropy increasing with
the radius $r$, while in the limit $\rai\to\infty$ the system becomes
globally isotropic.  As well known, the DF of the
density component $\roi$ is given by
\begin{eqnarray}\label{eq:fOM}
\fid(\qi)&=&\displaystyle\frac{1}{\sqrt{8}\pi^2}\frac{d}{d\qi}\int_0^{\qi}
\frac{d\roqi}{d\psit}\frac{d\psit}{\sqrt{\qi-\psit}}\nonumber \\
&=&\displaystyle\frac{1}{\sqrt{8}\pi^2}\int _0^{\qi} \frac{d^2\roqi}{
d\psit^2}\frac{d\psit}{\sqrt{\qi-\psit}}\mbox{,}
\end{eqnarray}
where the {\it augmented density} is
\begin{equation}\label{eq:rhoOM}
\roqi(r)\equiv\left (1+\frac{r^2}{\rai^2}\right) \roi (r),
\end{equation}
$\psit (r)=\sum_k\psii (r)$ is the total relative potential,
$0\leq\qi\leq\psit (0)$, and in the integrals above it is assumed that
the radius is eliminated from $\roqi$ in favour of $\psit$.  It can be
proved that the second equivalence in eq.~(\ref{eq:fOM}) holds for
untruncated systems with finite total mass, as the models discussed
here.

\subsection{Necessary and sufficient conditions for consistency}

Quite obviously, only a very limited number of density profiles among
those expressible in analytic form admit an explicit DF, so that the
study of phase--space consistency would appear restricted to such rare
cases when conducted analytically, while all the remaining cases
should be investigated numerically.  Fortunately this is not true, as
the CP92 technique is based on the verification (numerical or
analytical) of the following
\medskip
\parn
{\bf Theorem}: A {\it necessary condition} (NC) for the
non--negativity of $\fid$ is
\begin{equation}\label{eq:NC}
\frac{d\roqi(r)}{dr}\leq 0,\quad 0\leq r \leq\infty .
\end{equation}
If the NC is satisfied, a {\it strong sufficient condition} (SSC)
for the non--negativity of $\fid$ is
\begin{equation}\label{eq:SSC}
\frac{d}{dr}\left[\frac{d\roqi(r)}{dr} \frac{r^2\sqrt {\psit(r)}}{\mt
(r)}\right]\geq 0, \quad 0\leq r\leq\infty .
\end{equation}
Finally, a {\it weak sufficient condition} (WSC) for the non
negativity of $\fid$ is
\begin{equation}\label{eq:WSC}
\frac{d}{dr}\left[ \frac{d\roqi(r)}{dr}\frac{r^2}{\mt (r)}\right]\geq
0, \quad 0\leq r\leq\infty .
\end{equation}
\parn
{\bf Proof}: See CP92, C96, C99.
\parn
The first consideration that follows from the conditions above is that
the violation of the NC is connected only to the radial behavior of
the augmented density $\roqi$, and so this condition applies
independently of any other density component of the model.  Obviously,
a model failing the NC is {\it certainly inconsistent}, while a model
satisfying the NC {\it may be inconsistent}, i.e. the $\fid$ may be
negative even for values of model parameters allowed by the NC.  The
second consideration is that a model satisfying the WSC (or the more
restrictive SSC) is {\it certainly consistent}, while a model failing
the WSC (SSC) {\it may be consistent}. Therefore the consistency of an OM 
model satisfying the NC and failing the WSC (or the SSC) can be proved
only by direct inspection of its DF.

\subsubsection{A density slope--OM anisotropy inequality}

The NC can be recast into a simple inequality between the value of the 
density slope
\begin{equation}
\gamma(r)\equiv-\frac{d\ln\rho}{d\ln r}
\end{equation}
and the value of the orbital anisotropy indicator
\begin{equation}\label{betaOM}
\beta(r)\equiv 1-\frac{\sigt^2}{2\sigr^2}=\frac{r^2}{r^2+\ra^2}
\end{equation}
that must hold at each radius for consistent OM systems.
In the expression above $\sigr^2$ and $\sigt^2$ are the radial and
tangential components of the velocity dispersion tensor of the system
(e.g. Binney \& Tremaine 2008), and in the following proof we restrict
to natural systems, i.e. systems with monotonically decreasing
density profile, so that $\gamma(r)>0$. The proof is trivial: in fact, 
it suffices to express the NC in terms of the logarithmic density slope as
\begin{equation}
\frac{2}{\gamma(r)}-\frac{r^2}{r^2+\ra^2}\leq 0,
\end{equation}
and from eq.~(\ref{betaOM}) one
obtains the necessary condition
\begin{equation}\label{cuspslope}
\gamma(r)\geq 2\beta(r),
\end{equation}
which must be verified by each OM component of a consistent
multi--component system. Curiously, The condition above is formally
{\it identical} to the inequality appearing in the ``cusp
slope-central anisotropy'' theorem derived in AE06. This latter
theorem was proved by using constant anisotropy systems (i.e., $\beta
(r)=\beta$, and in this case as well the inequality above holds at all
radii in consistent models. See eq.~[10] in AE06), and then it was
convincingly argued that the inequality it holds asymptotically for
the central regions of spherical systems with generic anisotropy
distribution.

In the specific case of \ngamma density distributions
(see eq.~[\ref{eq:ng}]), the logarithmic density slope 
\begin{equation}\label{gammarhong}
\gamma(r)=\frac{\gamma+n s}{1+s},\qquad s={r\over\rc},
\end{equation}
and therefore eq.~(\ref{cuspslope}) shows that in general the NC is
satisfied near the center of OM systems (where $\beta=0$ and
$\gamma(r)=\gamma$), and at large radii (where $\beta =1$ and
$\gamma(r) = n\geq3$), while critical behaviors may be expected at
intermediate radii (see Sect.~4).

\subsection{Classification of phase-space inconsistency as a function of 
the anisotropy radius}

Interestingly (but not unexpectedly) the particular functional form of
the augmented density characteristic of OM parametrization, limits the
possible manifestations of phase--space inconsistency to few general
cases, that can be illustrated as follows. From
eqs.~(\ref{eq:fOM})-(\ref{eq:rhoOM}) it is apparent that the DF of
each density component can be written as
\begin{equation}
f(Q) =\fis(Q)+\frac{\fan(Q)}{\ra^2},
\label{eq:fSA}
\end{equation}
where $\fis$ is the DF of the considered density component in the
isotropic case (for simplicity, from now on we avoid the use of the
subscript $k$). Let $A_+$ be the subset of phase--space defined by
the property that $\fis$ is positive $\forall Q\in A_+$. Then, from
eq.~(\ref{eq:fSA}) a first condition to be satisfied for consistency
is
\begin{equation}\label{eq:sa-}
\ra\geq\racm \equiv\sqrt{\max\left\{0,{\rm sup}\left[-\frac{\fan(Q)}{
                     \fis(Q)}\right ]_{Q\in A_+}\right\}}.
\end{equation}
Obviously, when $\fis>0$ over all the accessible phase--space (the
common situation for realistic density distributions), inequality
(\ref{eq:sa-}) is also the {\it only} condition to be satisfied for
the model consistency, and there is at most a lower bound for the
anisotropy radius.  For example this is the case for the families of
one--component anisotropic $\gamma$ models (C99) and Sersic (1968)
models (Ciotti 1991).

When the complement of $A_+$ is not empty, i.e. $\fis <0$ over
some region $A_-$ of the accessible phase--space, a second inequality
again derived from eq.~(\ref{eq:fSA}) must be verified for
consistency:
\begin{equation}\label{eq:sa+}
\ra\leq\racp\equiv\sqrt{{\rm inf}\left[\frac{\fan (Q)}{
                       |\fis (Q)|}\right ]_{Q\in A_-}}.
\end{equation}
Therefore, if there exists some $Q\in A_-$ for which $\fan <0$, then
the proposed model is inconsistent\footnote{In C99 and Ciotti (2000)
  it is erroneously stated that the model is inconsistent if $\fan <0$
  $\forall Q\in A_-$.  All the results presented therein are however
  correct.}.  If $\fan$ is positive over $A_-$, from conditions
(\ref{eq:sa-})-(\ref{eq:sa+}) it follows that $\racm<\ra<\racp$ for OM
consistency, so that if $\racp<\racm$ the proposed model is
inconsistent.  Note that formally \textit{identical} considerations
hold when discussing the inequalities (\ref{eq:NC}) and
(\ref{eq:WSC}), because from eq.~(\ref{eq:rhoOM}) it follows that the
NC and WSC can be written in the same functional form of
eq.~(\ref{eq:fSA}): of course, in these cases the sets $A_+$ and $A_-$
are to be intended as subsets of the radial range $0\leq r\leq\infty$.

In the following Section, after presenting the one and two--component
\ngamma models, the necessary and sufficient conditions for model
consistency will be derived, also for the case of \ngamma models with
a massive central black hole.

\section {The one and two--component \ngamma models}

The one--component \ngamma models are a natural generalization of the
widely explored family of the so--called $\gamma$ models (Dehnen 1993,
Tremaine et al. 1994), whose density, cumulative mass within $r$, and
relative potential are given by
\begin{eqnarray}\label{eq:rhoG}
\rho(r)&=&\displaystyle\frac{3-\gamma}{4\pi}\frac{M\rc}
{r^{\gamma}(\rc+r)^{4-\gamma}},\\
M(r)&=&M\times\left(\frac{r}{\rc +r}\right)^{3-\gamma},
\end{eqnarray}
\begin{equation}\label{eq:PhiG}
\Psi(r)=\frac{GM}{\rc}
         \begin{cases}\displaystyle{\frac{1}{(2-\gamma)}
                \left [1-\left(\frac{r}{r+\rc}\right)^{2-\gamma}\right]}\cr
                \displaystyle{\ln\frac{r+\rc}{r}},\quad(\gamma=2),\cr
         \end{cases}
\end{equation}
respectively.

\begin{figure}
\includegraphics[height=0.46\textheight,width=0.52\textwidth]{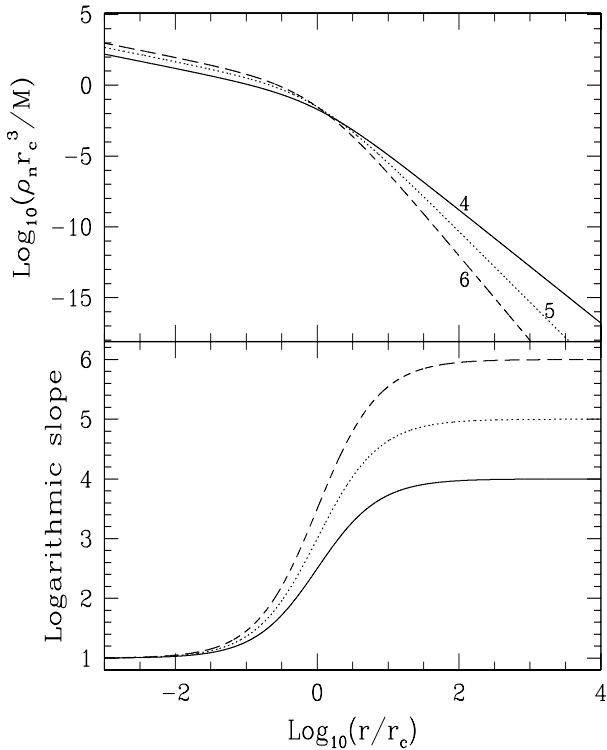}
\caption{The normalized density profile $\rho_n$ of the $n$-1 model
  (top panel), and its logarithmic slope (eq.~[\ref{gammarhong}], bottom panel), for
  $n=4,5$, and $6$.}
\label{f1}
\end{figure}

In the \ngamma models the logarithmic density slope for $r\gg\rc$ is
not 4, but it is a free parameter $n>3$ (to ensure that their total
mass is finite), so that these density profiles belong to the family
considered by Zhao (1996).  In the following, in order to exploit a
useful analytical property of the \ngamma density profiles, we will
often assume $n$ restricted to integer values $\geq4$.  In fact, the
generic density profile of a \ngamma model is given by
\begin{eqnarray}\label{eq:ng}
\rho_n(r)&\equiv&\frac{M\rc^{n-3}}
{4\pi\eulB(3-\gamma,n-3)r^{\gamma}(\rc+r)^{n-\gamma}}\nonumber\\&\\
&=&\frac{\rc^{n-3}(-1)^n}{\Gamma(n-3)}\frac{d^{n-4}}{d\rc^{n-4}}
\frac{\rho (r)}{\rc},\nonumber
\end{eqnarray}
where $M$ is the total mass of the density distribution,
$\eulB(x,y)=\Gamma(x)\Gamma(y)/\Gamma(x+y)$ and $\Gamma(x)$ are the
complete Euler beta and gamma functions, respectively, and the first
expression holds for any real number $n>3$.  The second expression,
based on repeated differentiation of eq.~(\ref{eq:rhoG}) with respect
to $\rc$, holds instead for integer $n\geq4$.  Of course, for $n=4$
the standard $\gamma$ model density profile is recovered.  The radial
behaviour of $\rho_n(r)$ is shown in Fig.~\ref{f1} (top panel) for 
$n$-1 models with increasing $n$.  In the bottom panel we show the
corresponding logarithmic density slopes, calculated accordingly to
eq.~(\ref{gammarhong}).  In particular, it is apparent how the inner slope is
$\gamma$ (for $r\lsim\gamma\rc/n$), and the external is $n$ (for
$r\gsim\rc$).  As a consequence, while the density profile outside
$\rc$ becomes more and more steep at increasing $n$, the innermost
region where the density slope is $\gamma$ contracts near the center.
As anticipated, the possibility to write the density $\rho_n$ as a
derivative with respect to the scale--length $\rc$ provides an easy
way to determine analytical properties of the \ngamma models for $n$
integer. Indeed, since the density enters linearly in the
integrals of the total mass inside the radius $r$ and of the
potential, from eq.~(\ref{eq:ng}) it follows that
\begin{eqnarray}\label{eq:Mng}
M_n(r)&=&\frac{\rc^{n-3}(-1)^n}{\Gamma(n-3)}\frac{d^{n-4}}
{d\rc^{n-4}}\frac{M(r)}{\rc},\\
\Psi_n(r)&=&\frac{\rc^{n-3}(-1)^n}{\Gamma(n-3)}
\frac{d^{n-4}}{d\rc^{n-4}}\frac{\Psi(r)}{\rc}.\label{eq:Phing}
\end{eqnarray}
Instead, when $n$ is not at integer, $M_n$ and $\Psi_n$ are in general
given by hypergeometric ${}_2F_1$ functions. Expressions similar to
eqs.~(\ref{eq:Mng})-(\ref{eq:Phing}) hold for any quantity that can be
written as a linear functional of $\rho_n$, so that the surface
density profile of \ngamma models with $n$ integer can be obtained by
repeated differentiation of the surface density profile of $\gamma$
models (when analytically available, e.g. see Binney \& Merrifield
1998).  This property will be exploited in Section 4 to obtain the
explicit DF of halo dominated (\ngam,$\gau$) models; moreover, in
Appendix B we show how quadratic functionals of $\rho_n$ (such as the
gravitational energy and the velocity dispersions) can be also
evaluated by using repeated differentiation with respect to $\rc$.

We are now in the position to introduce the two--component models used
in this work.  The most general family of two--component \ngamma
models, i.e. the ($n_1$-$\gau$,$n_2$-$\gad$) models, is made of the
superposition of two \ngamma models with different total masses,
scale--lengths, internal and external density slopes, and finally two
different anisotropy radii. For simplicity here we restrict to the
case of (\ngam,$\gau$) models, where the ``halo'' density distribution
is a standard $\gau$ model: we note however that some of the presented
results can be generalized without much effort to the family of
($n_1$-$\gau$,$n_2$-$\gad$) models (see Appendix B).  In the
following, the total mass $\ms$ and the characteristic scale--length
$\rs$ of the $\gau$ halo are adopted as normalization constants, so
that the physical scales for density and potential are given by
$\rhon=\ms/\rs^3$ and $\psin =G\ms/\rs$, while we define $s\equiv
r/\rs$, $\xi\equiv\rc/\rs$, $\mu\equiv M/\ms$.  With this choice, the
(\ngam,$\gau$) models are structurally determined by fixing the four
independent parameters $(\ms,\rs,\mu,\xi)$, with the obvious
conditions $\mu\geq 0$ and $\xi\geq 0$. We conclude this introductory
discussion by noticing that, for reasons that will become apparent in
Section 3.2 and 4, the present normalization differs from that adopted
in C99, where the normalization mass and scale--length were those of
the first component.

\subsection{The necessary and sufficient conditions for one--component
  \ngamma models}

Before discussing the case of the two--component models, we consider
the NC for anisotropic \ngamma models, in order to determine
analytically a critical value for the anisotropy radius such that a
higher degree of radial OM anisotropy (i.e., a smaller $\ra$) would
produce a negative DF for some permitted value of $Q$.  We recall that
the obtained anisotropy limit holds also when a second component is
added (see Section 2), so that the present discussion is fully
general. Moreover, we note that as the NC involves only the density
distribution under scrutiny, we can use the first expression in
eq.~(\ref{eq:ng}), and the following results hold for any $n>3$, not
necessarily limited to integers. In the following the unit mass and
unit length are the total mass $M$ and the scale--length $\rc$ of the
\ngamma model, with $\sa=\ra/\rc$.

In C99 the analytical expression for the critical $\sa(\gamma)$ was
obtained for the whole family of $\gamma$-models, and here we derive
its generalization $\sa(\gamma,n)$. In fact, eq.~(A1) shows that for
$2\leq\gamma <3$ and $n>3$ the NC is satisfied for $\sa\geq 0$, and
the result of C99 is now obtained as the particular case for $n=4$.
In other words, the NC leaves open the possibility of making \ngamma
models with $\gamma\geq 2$ using radial orbits only. In the range
$0\leq \gamma <2$ the NC requires instead that
\begin{equation}
\sa\geq \sM\sqrt{\frac{2-\gamma-(n-2)\sM}{\gamma+n\sM }},
\label{eq:ngNC}
\end{equation}
where the explicit expression of $\sM(n,\gamma)$ is given by eq.~(A2),
and the inequality above reduces to eq.~(13) of C99 when $n=4$. The
NC then proves that \ngamma models with $0\leq \gamma <2$ cannot be
made of radial orbits only, independently of the value of the external
slope $n$, and of the presence of any possible second component.  This
result extends the list of cases for which it has been proved that a
density cusp shallower than $r^{-2}$ cannot be supported by radial
orbits only (Richstone \& Tremaine 1984, CP92, C99, AE06).
\begin{figure}
\includegraphics[height=0.4\textheight,width=0.5\textwidth]{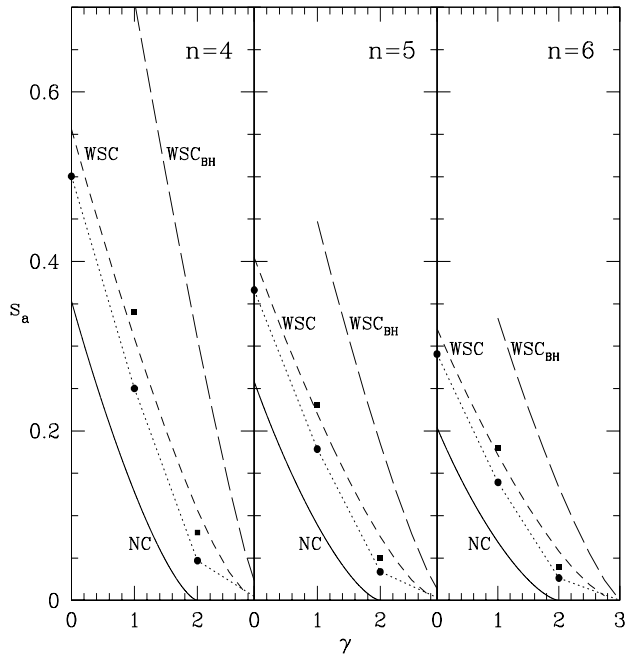}
\caption{Consistency limits on $\sa=\ra/\rc$ as a function of $\gamma$
  in one--component \ngamma models, for $n=4,5,6$.  The NC limit is
  represented by the solid lines: all models with the pair
  ($\sa$,$\gamma$) in the triangular region below them are
  inconsistent.  The dashed lines mark the WSC limit: all points above
  these lines correspond to consistent one--component \ngamma
  models. For $\gamma=0,1,2$, the filled circles joined by the dotted
  line represent the more accurate limits obtained from the SSC. For
  the black-hole dominated ($n$-$\gamma$,{\rm BH}) models the long-dashed
  lines (interrupted for $\gamma<1$, see Section 3.2), are the lower
  limit on $\sa$ given by the WSC, while the solid squares are the
  true limits derived from the DF.}
\label{f2}
\end{figure}
In Fig.~\ref{f2} the lower bound for the anisotropy radius given in
eq.~(\ref{eq:ngNC}) is shown with the solid lines as a function of
$\gamma$ for $n=4,5,6$: \ngamma models (one or multi--component) with
the pair ($\sa$,$\gamma$) in the nearly triangular region under the
solid curves are inconsistent.  In particular, for fixed $\gamma$ an
increase of $n$ produces a decrease of the minimum $\sa$, i.e. 
we have here a first indication that a steepening of the density profile
in the external regions of a system with fixed inner density slope can
be effective in increasing the maximum amount of sustainable radial
anisotropy.  This behaviour is quantified by substitution in
eq.~(\ref{eq:ngNC}) of the asymptotic expansion for $n\to\infty$ of
$\sM(n,\gamma)$ in eq.~(A2):
\begin{equation}
\sM(n,\gamma)=\frac{1-2\gamma+\sqrt{1+4\gamma}}{2n}+{\rm O}(n^{-2}).
\label{eq:smAs}
\end{equation}
As often happens in asymptotic analysis, even if the expansion above
holds in principle only for very large values of $n$, the substitution
of (\ref{eq:smAs}) in (\ref{eq:ngNC}) leads to percentual errors on
$\sa$ less than 22\%, 14\%, and 10\% for $n=4,5,6$ respectively (for
the inner density slope $\gamma=1$).

We now move to discuss the WSC for one--component \ngamma models: the
obtained $\sa(n,\gamma)$ will mark a lower limit above which
consistency (for the considered $n$ and $\gamma$) is guaranteed.
Unfortunately, the WSC cannot be explored algebraically in the general
case, because the resulting inequality (that for simplicity we do not
report here) involves the solution of an equation of fifth degree for
$n=5$, and the degree increases for increasing $n$.  For $n=4$,
instead, it is possible to treat the WSC analytically, since it
reduces to the discussion of a cubic equation (see C99).  Of course,
the critical values for the anisotropy radius can be easily obtained
solving numerically inequality (\ref{eq:WSC}) in any specific case of
interest, and the results are shown in Fig.~\ref{f2} for $n=4,5,6$
with dashed lines: all one--component \ngamma models with the pair
($\sa$,$\gamma$) in the region above the dashed lines are consistent.
Note how for increasing values of the external density slope $n$ more
and more radially anisotropic orbital distributions can be supported,
thus confirming the indications provided by the NC.

Values of $\sa$ nearer to the limits set by the DF are obtained by
using the SSC.  Inequality (\ref{eq:SSC}) evaluated for a generic pair
($\gamma$,$n$) results in a transcendental equation that must be
studied numerically (as already done for $n=4$ in C99), and the black
dots joined by the dotted lines in Fig.~\ref{f2} represent the
critical lower values of $\sa$ for one--component $\gamma$ models with
$\gamma=0,1,2$, and $n=4,5,6$.  As expected, the dotted lines are
contained between the solid (NC) and the dashed (WSC) lines in each
panel, and again they shift downward for increasing $n$.

Thus, from this preliminary investigation of one--component \ngamma
models we conclude that for fixed $\gamma$ an increase of $n$
corresponds to a decrease of the minimum admissible value of $\sa$,
i.e. \textit{steeper density distributions in the external regions can
  support more radial anisotropy}.

\subsection{Sufficient conditions for halo--dominated (\ngam,$\gau$) models}

In order to proceed further with the preliminary discussion, we now
apply the WSC to the \ngamma component of a (\ngam,$\gau$) model,
extending to this class of systems some of the results obtained in C96
and C99 for two--component $(\gau,\gad)$ models.  In particular, the
analytical study in C99 was restricted to some representative
$(\gau,\gad)$ models, namely $a)$ isotropic two-component systems with
inner slopes in the ranges $1\leq\gau<3$, and $0\leq\gad\leq\gau$
(i.e. the $\gad$ component is shallower in the central regions); $b)$
isotropic two--component (0,1) systems (i.e. the $\gad$ component is
steeper in the central regions); and finally $c)$ anisotropic $\gau$
profiles with $1\leq\gau<3$ in the gravitational field of a dominant
central black hole.  Here, in order to obtain analytical results for
the more general ($\ngam$,$\gau$) models, we assume that the $\gau$
component (the ``halo'') dominates everywhere the gravitational field.
Under this simplifying assumption, the following three results,
corresponding to the points $a)$, $b)$, and $c)$ above, will be proved
analytically:
\begin{enumerate}
\item In the case of halo--dominated {\it isotropic} (\ngam,$\gau$) models,
  with $1\leq\gamma <3$, $0\leq\gau\leq\gamma$, and $n>3$, the
  centrally more peaked \ngamma component is consistent, independently
  of the value $\xi=\rc/\rs$ of its concentration relative to the
  centrally flatter $\gau$ halo.  In the case of {\it anisotropic}
  (\ngam,$\gau$) models, the determination of a minimum anisotropy
  radius for consistency as a function of $n$,$\gamma$,$\gau$,$\xi$
  reduces to the solution of an algebraic equation of sixth degree
  (for generic $n$), which is solved numerically.  In the particular
  case of halo--dominated ($n$-2,1) models, the application of the WSC
  shows that for $\xi\leq (n-1)/2$ these models can be consistently
  assembled using radial orbits only.

\item In the case of halo--dominated {\it isotropic} ($n$-0,1) models, the
  WSC shows that for $\xi\leq (n+1)/2$ the $n$-0 density distribution
  is consistent.  For broader $n$-0 density distributions, the models
  can be consistent only in presence of some amount of radial
  anisotropy.

\item In the case of {\it anisotropic} $\ngam$ models with a dominant black
  hole at their center it is possible to determine analytically a
  lower limit $\ra(n,\gamma)$ for consistency, and this limit
  decreases for increasing $n$.
\end{enumerate}
\begin{figure}
\includegraphics[height=0.4\textheight,width=0.5\textwidth]{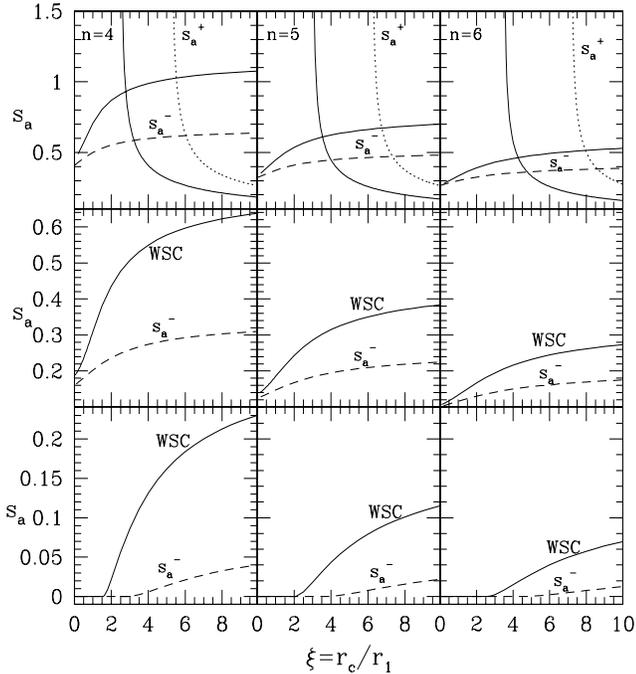}
\caption{Minimum value of the anisotropy radius (normalized to the
  scale--length of the \ngamma component, i.e. $\sa=\ra/\rc$) as a
  function of the relative concentration $\xi$, for consistency of
  halo dominated ($\ngam$,1) models.  The external density slope $n$
  increases as $n=4,5$, and $6$ from left to right panels, while the
  internal density slope as $\gamma=0,1$, and $2$ from top to bottom
  panels.  Solid lines represent the limits derived from the WSC,
  while dashed lines are the $\sacm$ derived from the DF.  Note that
  for ($n$-0,1) models also the $\sacp$ limits appear for
  $\xi>\xi_{\rm{c}}$ (see Sections 3.2 and 4.3 for details).}
\label{f3}
\end{figure}
The proof of the first result, which is an extension of the study
mentioned in point $a)$, is conceptually straightforward but
algebraically cumbersome, and only the main steps are reported in
Appendix A (eq.~[A3]). In particular, as this result holds also for
$n>3$, while the external logarithmic slope of the $\gau$ halo is $4$,
it means that a centrally steep density profile can be consistent in
the gravitational field of a more massive (but centrally shallower)
component, even when its external regions are less peaked.  The
situation is illustrated for some representative cases in
Fig.~\ref{f3} (bottom and middle panels), where all points above the
solid lines correspond to consistent halo--dominated ($n$-1,1) and
($n$-2,1) models, and apparently the isotropic limit ($\sa=\infty$) is
allowed for any choice of $\xi$; the solid lines are obtained by
solving numerically the corresponding WSC. A few additional trends are
apparent. First, for a given central logarithmic density slope
$\gamma$ and for a given relative concentration $\xi$, an increase of
$n$ corresponds to a better and better ability to sustain radial
anisotropy, and thus the trend found in one--component models is
confirmed also in the two--component case. Second, at fixed $n$ and
$\gamma$, the minimum anisotropy radius for consistency increases for
increasing $\xi$. This trend was already found in C96 and C99: in
practice, in a fixed potential broader density distributions are less
and less able to sustain radial anisotropy. Third, at fixed external
slope $n$ and relative concentration $\xi$, more centrally peaked
systems are better able to support radial anisotropy. An additional
comment concerns the specific case of ($n$-2,1) models (Fig.~\ref{f3},
bottom panels). In fact, it is apparent how in presence of a dominant
$\gau=1$ halo the WSC limits flatten to zero for relative
concentration less than some critical value $\xi_{\rm{c}}$, and
accordingly the model may be purely radially anisotropic.  Remarkably,
it is easy to show that the $n$-2 component can support purely radial
orbits for relative concentrations $\xi\leq\xi_{\rm{c}}=(n-1)/2$ (see
eq.~[A4]), which are exactly the critical points in Fig.~\ref{f3}. Of
course, this is just a sufficient condition, so we expect that the
existence of a larger critical concentration for consistency in case
of purely radial orbits will be revealed by direct inspection of the
DF.  A final comment is due. In fact, the adopted range of values
  for the internal density slopes $\gamma\geq\gamma_1$ imply that the
  halo density distribution is centrally less peaked than the
  $\ngamma$ component. Therefore, for large but {\it finite} total
  halo mass, the integrated mass of the halo is subdominant with
  respect to the stellar one for vanishing $r$, and the assumption of
  a dominant halo breaks down at the very center.  This means that in
  an asymptotic sense, the analysis presented in the previous Section
  should be applied at the center of these finite-halo mass models.

Result (ii) extends the study mentioned in point $b)$ above, and
it is proved in eq.~(A5).  The obtained limitation
$\xi\leq\xi_{\rm{c}}=(n+1)/2$, shown by the solid line in Fig.~\ref{f4},
suggests that the concentration of the $\gamma=0$ component must
``adapt'' to the dominant $\gau=1$ halo for phase--space consistency.
Note that $\xi_{\rm{c}}$ increases for increasing $n$, i.e.  a steeper
external density profile can (partially) compensate for the effect of
a central shallower density distribution.  The appearence of the
concentration limit $\xi_{\rm{c}}$ in the isotropic case manifests in
the top panels of Fig.~\ref{f3}.  In fact, as discussed in Section 2,
in systems with positive isotropic functions (WSC or DF) only a lower
limit for $\sa$ exists, as in the case of halo dominated ($n$-1,1) and
($n$-2,1) models.  In the present case, instead, when for increasing
$\xi$ the critical value $\xi_{\rm{c}}=2.5$ is crossed (in the $n=4$
panel), the additional condition $\sa<\sacp$ corresponding to the
radial domain $A_-$ appears as the vertical asymptote of the solid
line, and the parameter space for consistency progressively reduces,
shrinking to zero for the value of $\xi$ where the two solid lines
(obtained numerically) cross.  A similar configuration repeats in the
panels representing the $n=5$ and $n=6$ cases, where the consistency
limits move, as expected, downward and rightward.  Of course, these
limits represent only a sufficient condition for the consistency, and
when considering the DF in the limit $\mu\to 0$ (see Section 4) we
expect to determine larger critical values of $\xi_{\rm{c}}$, and
larger consistency regions in the ($\xi$,$\sa$) plane.
 
The result concerning the models with central black hole, related
to point $c)$ above, is proved as follows.  As well known, for
$\rs\to 0$ the potential of the $\gau=1$ model becomes that of a point
mass, and from the first result proven in this Section it follows that
a black hole of any mass $M_{\rm BH}$ can be consistently added at the
center of a globally isotropic \ngamma model, when $1\leq\gamma<3$ and
$n>3$.  For this reason in the following we restrict to this range of
slopes $\gamma$, and accordingly the long--dashed curves in
Fig.~\ref{f2} interrupt for $\gamma<1$.  Having reduced the study to
models with a positive isotropic DF, we only have to estimate a lower
limit of $\ra$.  In Appendix A the WSC is applied to the anisotropic
(\ngam,BH) models with $1\leq\gamma <3$; it can be discussed
analytically in the special case of a {\it dominant} black hole, i.e.
assuming in eq.~(\ref{eq:WSC}) $\mt =M_{\rm BH}$ (and so
$\psit=GM_{\rm BH}/r$).  As shown in eq.~(A6), under these assumptions
\begin{equation}\label{eq:saBH}
\sa\geq\sM
\sqrt\frac{(3-\gamma)(\gamma-2)+\sM(n-2)[(6-2\gamma)-(n-3)\sM]}{
      n(n-1)\sM^2+2n(\gamma-1)\sM+\gamma(\gamma-1)},
\end{equation}
where $\sM=\sM(n,\gamma)$ is obtained by solving an algebraic equation
of fourth degree: the obtained limits are shown in Fig.~\ref{f2} with
the long--dashed lines, for $n=4,5,6$.  For $n=4$ equation above
coincide with that given in eq.~(18) of C99, while asymptotic analysis
proves that for $n\to\infty$
\begin{equation}\label{eq:smBHas}
\sM(n,\gamma)=\frac{s_{\rm M 0}(\gamma)}{n}+{\rm O}(n^{-2}),
\end{equation}
where $s_{\rm M0}(\gamma)$ is the larger real root of eq.~(A9).  We
found that the substitution of eq.~(\ref{eq:smBHas}) in
eq.~(\ref{eq:saBH}) leads to estimates of the minimum $\sa$ discrepant
from the values obtained by the full formula by less than 50\%, 33\%,
and 23\% for $n=4,5,6$ respectively (for the inner density slope
$\gamma=1$).  As in the other cases, 
an increase of the external density slope $n$ 
makes it possible to sustain more radial orbits.

\section{The DF of halo-dominated (\ngam,$\gau$) models}

It should be clear that the DF of two--component (\ngam,$\gau$) models
cannot be obtained analytically, except for very special combinations
of the values of $n,\gamma,$ and $\gau$, as for example the (1,1)
models discussed in C96 (that would be referred as ($4$-$1$,$1$)
models in the present notation).  However, the search is not hopeless.
In fact, from C99 it is known that the halo dominated DF of
two--component (1,0) and (0,1) galaxy models can be obtained in terms
of elementary functions, and a simple argument shows that this is also
the case of (2,1) and (2,0) models.  This fact, the linearity of the
OM inversion formula with respect to $\rho$ in the non
self--gravitating case (see eq.~[2]), and finally the 
formula (\ref{eq:ng}) for $n$ integer, prove that also the DF of the
halo dominated (\ngam,1) and (\ngam,0) models can be expressed in
terms of elementary functions, for any integer $n\geq4$ and
$\gamma=0,1,2$.  Note that, albeit their special nature, the study of
halo--dominated models is useful because the formulae -- expressible
using elementary functions -- can be studied very easily, making
apparent the effects of the relative distribution of the investigated
component and of the halo.  In particular, in this Section we will
determine the DF of the \ngamma component of halo dominated (\ngam,1)
models, and the exact phase--space constraints will be derived and
compared with those obtained using the NC, WSC, and SSC in Section 3.
We restrict to the $\gau=1$ case because the Hernquist (1990)
potential is the simplest in the class of $\gamma$ models;
furthermore, this choice allows us to investigate the consistency of
models with central density slopes flatter ($\gamma=0$), equal
($\gamma=1$), and steeper ($\gamma=2$) than that of the halo.  As a
consequence, the results obtained for (\ngam,1) models should be
representative of the whole situation.

With the normalization scales introduced at the beginning
of Section 3, the relative potential of the Hernquist halo
needed for the recovery of the DF is
\begin{equation}\label{eq:Her}
\Psi(r) =\displaystyle{\frac{\psin}{1+s}}=\psin \psitil(s),
\end{equation}
while from eq.~(\ref{eq:ng})
\begin{equation}\label{eq:rhoNor}
\rho_n (r)=\rhon \frac{\mu\xi^{n-3}(3-\gamma)(-1)^n}{4\pi\Gamma(n-3)s^{\gamma}}
                 \frac{d^{n-4}(\xi+s)^{\gamma-4}}{d\xi^{n-4}}.
\end{equation}
As for the density and the potential, also for the DF it is useful to
work with dimensionless functions, and we define $f=\fn\ftil(\Qtil)$
with $\fn=\rhon\psin^{-3/2}$, and $0\leq\Qtil=Q/\psin\leq1$: note
that, at variance with C99, the normalization quantities used here are
those of the halo, consistently with the halo dominated nature of the
present models. In other words, the dominant density distribution sets
the natural scales.  The easiest way to compute the DF is to change
the integration variable from the total potential to the radius in the
first of the identities in eq.~(\ref{eq:fOM}).  After normalization we
obtain
\begin{equation}\label{eq:fOMnor}
f(Q)=\frac{\fn}{\sqrt{8}\pi^2}\left(\frac{d\Qtil}{d\nu}\right)^{-1}
\frac{d\Ftil(\nu)}{d\nu},\quad \nu=\nu (\Qtil),
\end{equation}
where
\begin{eqnarray}\label{eq:fnu}
\Ftil(\nu)&=&-\int_{\nu}^{\infty}\displaystyle\frac{d\tilde\varrho_n}{ds}
\frac{ds}{\sqrt{\psitil(\nu)-\psitil(s)}}\nonumber \\
&=&\Ftilis(\nu)+\frac{\Ftilan (\nu)}{\sa^2},
\end{eqnarray}
and from eq.~(\ref{eq:Her})
\begin{equation}\label{eq:Qnu}
\Qtil=\frac{1}{1+\nu},\quad \left(\frac{d\Qtil}{
d\nu}\right)^{-1}=-(1+\nu)^2, \qquad 0\leq \nu\leq\infty.
\end{equation}
In the equations above, a negative sign appears in front
of the integral (\ref{eq:fnu}) due to the monotonic decrease
of the relative potential with increasing radius;
$\tilde\varrho_n$ is the normalized augmented density
associated with eq.~(\ref{eq:rhoNor}), while $\sa =\ra/\rs$ is the
normalized anisotropy radius; finally, the subscripts of the functions $F$
refer to the isotropic and anisotropic parts of the DF, respectively.
It is now evident how in halo dominated ($\ngam$,$\gau$) models the
external slope parameter appears in the density $\tilde\varrho_n$
only, so that the DF of the \ngamma component can be obtained by
repeated differentiation with respect to $\xi$ of the simpler DF of
the halo dominated ($\gamma$,$\gau$) model.  In the following, we will
also discuss black hole dominated ($\ngam$,BH) models: then, we will adopt
as natural normalization scales the black hole mass $M_{\rm BH}$ and
the scale--length $\rc$ of the \ngamma component; moreover,
$0\leq\Qtil\leq\infty$, since in this case $\psitil=1/s$, 
so that in eq.~(\ref{eq:Qnu}) $\Qtil=1/\nu$.

It is not difficult to show that, for generic values of $\gamma$,
the integral (\ref{eq:fnu}) can be expressed as a combination of
hypergeometric ${}_2F_1$ functions when the halo is a $\gau=0$, or
$1$ model, or a black hole.  However, the resulting expressions are
not more illuminating than the integral itself, so that we do not
report them here.  Instead, we prefer to show simple asymptotic
expansions relative to a couple of interesting situations.  The first
concerns the behaviour of the DF of ($\ngam$,BH) models for
$\Qtil\to\infty$, for which the leading term of the expansion
(normalized to $\fn/\sqrt{8}\pi^2$) is
\begin{equation}\label{asymBH}
\ftil(\Qtil)\sim\gamma\,\left(\gamma-\frac{1}{2}\right)\Qtil^{\gamma-3/2}:
\end{equation}
thus, the slope $\gamma=3/2$ marks the different behaviour of the
models. We note also that $f<0$ for $\gamma<1/2$ and the model is
inconsistent (Tremaine et al. 1994); furthermore, we note how 
in the leading term above the function $\Ftilan$ is producing no contribution.  
The other case of interest is the behaviour of the DF of isotropic \ngamma
models for $\Qtil\to 0$.  We recall that this case is very general, as
it applies to self-gravitating \ngamma models but also to \ngamma
models in generic halos of finite total mass, because in all systems
$\psit\sim G M_{\rm T}/s$ for $s\to\infty$. In this case we found
\begin{equation}
\ftil(\Qtil)\sim n\,\left(n-\frac{1}{2}\right)\Qtil^{n-3/2},
\end{equation}
an expression (obviously) formally identical to eq.~(\ref{asymBH}).
Finally, we notice that the asymptotic expansion for $r\to\infty$ of the
velocity dispersion of isotropic \ngamma models (both single component and
embedded in dark matter halos of finite total mass) is $\sigr^2\sim
1/(n+1)s$, normalized to $GM_{\rm{T}}/r_{\rm c}$, so that for
$r\to\infty$ $\rho/\sigr^3\sim (n+1)^{-3/2} s^{3/2-n}$, in accordance
with the comment in Footnote 1.

\subsection{The halo dominated ($n$-2,1) Model}

Following the preliminary discussion, here we derive the explicit
expression for the DF of the $n$-2 model with an arbitrary degree of
OM orbital anisotropy, in the gravitational field of a dominant
Hernquist halo.  After partial fraction decomposition of
eq.~(\ref{eq:fnu}), the isotropic and anisotropic components of the DF
are
\begin{equation}\label{dfisg2}
\Ftilis(\nu)=\displaystyle\frac{(-1)^{n}\mu\sqrt{1+\nu}\,\xi^{n-3}}{2\pi(n-4)!}\,
\,\displaystyle\frac{d^{n-4}}{d\xi^{n-4}}\displaystyle
\left(\frac{G_3+G^0_3}{\xi^2}+\frac{G_2-G^0_2}{\xi^3}\right),
\end{equation}
and
\begin{eqnarray}\label{dfang2}
\Ftilan(\nu)&=&\displaystyle\frac{(-1)^{n}\mu\sqrt{1+\nu}\,\xi^{n-3}}
                             {2\pi(n-4)!}\,
\displaystyle\frac{d^{n-4}G_3}{d\xi^{n-4}}\nonumber\\
&=&\displaystyle\frac{\mu\sqrt{1+\nu}(n-2)!\,\xi^{n-3}\,G_{n-1}}{4\pi(n-4)!},
\end{eqnarray}
respectively, where for $k\geq2$ we define
\begin{equation}
G_k(\xi,\nu)\equiv\int_{\nu}^{\infty}\sqrt{\frac{s+1}{s-\nu}}
                 \frac{ds}{(\xi+s)^k}, \qquad G^0_k(\nu)\equiv G_k(0,\nu).
\end{equation}
For $\xi=1$ the function $G_k$ assume a very simple form:
\begin{equation}
G_k(1,\nu)=\frac{\sqrt{\pi}\,\Gamma (k-1)}{\Gamma(k-1/2)
(1+\nu)^{k-1}},
\end{equation}
while for $\xi\neq 1$ the formula
\begin{equation}\label{recG}
G_{k+1}(\xi,\nu)=-\frac{1}{k}\frac{dG_k(\xi,\nu)}{d\xi}=
\frac{(-1)^{k-1}}{k!}\frac{d^{k-1}G_2(\xi,\nu)}{d\xi^{k-1}}
\end{equation}
holds, with
\begin{eqnarray}
&G_2(\xi,\nu)=\displaystyle{\frac{1}{\xi+\nu}}&\nonumber\\
&+\displaystyle{\frac{1+\nu}{
(\xi+\nu)^{3/2}}}&\begin{cases}
               \displaystyle{\frac{1}{\sqrt{1-\xi}}
               \arctan \sqrt{\frac{1-\xi}{\xi +\nu}}},
               &(0\leq\xi <1),\cr
               \displaystyle{\frac{1}{\sqrt{\xi-1}}
               \arctanh \sqrt{\frac{\xi -1}{\xi +\nu}}},
               &(\xi >1).\cr
                \end{cases}
\end{eqnarray}
Note that the second expression in eq.~(\ref{dfang2})
has been obtained from the relation (\ref{recG}).
\begin{figure}
\includegraphics[height=0.30\textheight,width=0.4\textwidth]{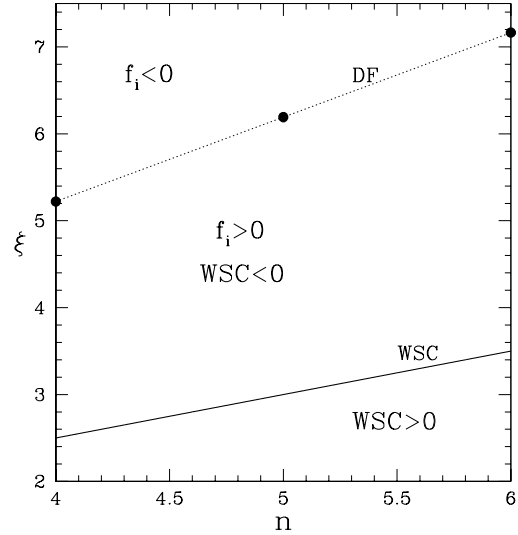}
\caption{Upper limits for consistency on the concentration parameter
  $\xi=\rc/\rs$ of the fully isotropic halo-dominated ($n$-0,1)
  model, as a function of the external density slope $n$.  The solid line is
  the limit derived from the WSC (Section~3.2), while the dotted line is the true
  limit obtained from the DF (Section~4.3).}
\label{f4}
\end{figure}
\begin{figure}
\includegraphics[height=0.45\textheight,width=0.55\textwidth]{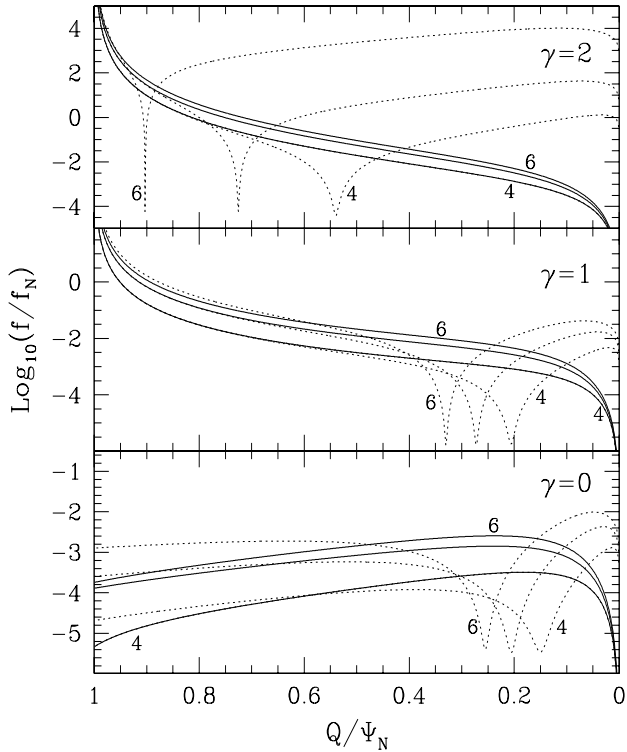}
\caption{From top to bottom: the dimensionless DF for the halo
  dominated ($n$-2,1), ($n$-1,1), and ($n$-0,1) models, as a function
  of $\Qtil$.  The solid lines represent the isotropic DFs, while the
  dotted lines are the anisotropic DFs, with $\sa$ approaching the
  critical value for consistency. In all cases, $\xi=5$.}
\label{f5}
\end{figure}
From the derived formulae we determined the true anisotropy limit for
the ($n$-2,1) models. We found that $\fis>0$ for all the explored
values of $n$, i.e. the isotropic $n$-2 model in a dominant Hernquist
halo is consistent independently of the halo scale--length, in
accordance with the result in point $1)$ of Section 3.2, obtained with
the WSC. Therefore the set $A_+$ in phase--space coincides with the
whole accessible phase--space, and only the anisotropy limit $\sacm$
exists. In Fig.~\ref{f5} (top panel) we show the isotropic (solid) and
strongly anisotropic (i.e., near to the consistency limit) DFs for
$n=(4,5,6)$ and $\xi=5$.  Note how the isotropic DFs are very similar,
and monotonically decreasing.  Instead, in the anisotropic cases the
depression leading to inconsistency is apparent, clearly showing how
phase--space inconsistency is always set outside the center.  This
feature is qualitatively similar to the others explored OM systems
(e.g., see Fig.~2 in Ciotti \& Lanzoni 1997, Figs.~2 and 3 in C99,
Fig.~3 in Ciotti, Morganti \& de Zeeuw 2008); in addition, the
systematic shift of the DF depression towards high $Q$ values for
increasing $n$ is also apparent: an argument supporting this
phenomenon is given in Section~2.1.1. We notice that the cuspy
  dips shown by the DFs for values of $\sa$ near the consistency limit
  could be the source of kinetic instabilities, whose investigation is
  of course well beyond the framework of this paper. In any case, it
  is almost certain that any system would develop radial orbit
  instability for values of the anisotropy radius larger than the
  consistency limit (e.g., see the N-body experiments discussed in
  Nipoti, Londrillo \& Ciotti 2002): therefore, the critical DFs in
  Fig.~5 describe equilibrium unstable systems.

The trend of the anisotropy limit $\sacm$ for different values of $n$
and increasing $\xi$ is shown in Fig.~\ref{f3} (bottom panel) with the
dashed lines. First, for fixed $n$ and increasing $\xi$, $\sacm$
increases: in practice, for a given halo, a broader stellar density
distribution is less and less able to sustain radial anisotropy.
Second, at fixed $\xi$ a steeper external density slope (i.e. larger
$n$) corresponds to higher amount of admissible radial anisotropy,
thus showing that not only the inner density slope, but also the
external density profile is important for phase--space consistency.
As expected, the true limitation on $\sacm$ is less stringent than the
corresponding limit derived from the WSC, and the dashed line is
everywhere below the solid line. In particular, while from the WSC the
critical concentrations $\xi_{\rm{c}}$ in order to have a purely
radial model are $1.5,2,$ and $2.5$ (for $n=4,5,$ and $6$), from the
DF we obtain $\xi_{\rm{c}}\simeq 2.8,3.7,$ and $4.5$.

It is also possible to determine the DF of ($n$-2,BH) models in
presence of a dominant central black hole. The resulting formulae
  are identical to eqs.~(\ref{dfisg2})-(\ref{dfang2}), where the
  quantity $\sqrt{1+\nu}$ at numerator is replaced by $\sqrt{\nu}$ and
  the $G$ functions are replaced by the corresponding $G_{\bullet}$
  functions of same index, defined for $k\geq 2$ as
\begin{equation}
G_{\bullet k}(\xi,\nu)\equiv\int_{\nu}^{\infty}\sqrt{\frac{s}{s-\nu}}
                 \frac{ds}{(\xi+s)^k},
\end{equation}
and
\begin{equation}
G^0_{\bullet k}(\nu)\equiv G_{\bullet k}(0,\nu)=\frac{\sqrt{\pi}\,\Gamma (k-1)}
                                                  {\Gamma(k-1/2)\nu^{k-1}}.
\end{equation}
For $\xi>0$ the formula
\begin{equation}
G_{\bullet k+1}(\xi,\nu)=-\frac{1}{k}\frac{d G_{\bullet k}(\xi,\nu)}{d\xi}=
\frac{(-1)^{k-1}}{k!}\frac{d^{k-1}G_{\bullet 2}(\xi,\nu)}{d\xi^{k-1}}
\end{equation}
holds, where
\begin{equation}
G_{\bullet 2}(\xi,\nu)=
\frac{1}{\xi+\nu}+\frac{\nu}{(\xi+\nu)^{3/2}\sqrt{\xi}}
\arctanh{\sqrt{\frac{\xi}{\xi+\nu}}}.
\end{equation}
\begin{figure}
\includegraphics[height=0.46\textheight,width=0.52\textwidth]{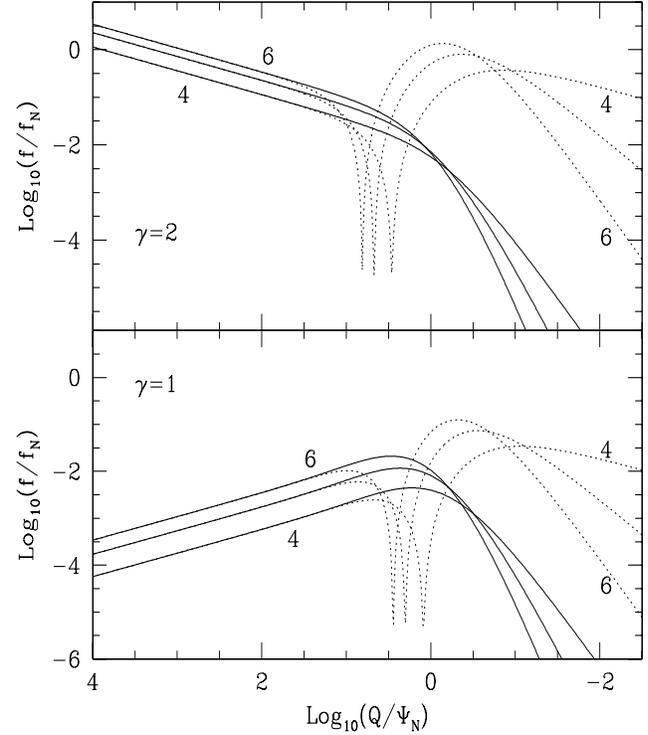}
\caption{Dimensionless DF for the black hole dominated ($n$-2,1) and
  ($n$-1,1) models, as a function of the relative energy $\Qtil$, for
  $n=4,5,$ and $6$. Normalization scales are the mass of the black
  hole and the scale--length of the \ngamma component.  Solid lines
  represent the isotropic DFs, while dotted lines the anisotropic DFs,
  with $\sa$ approaching the critical value for consistency.}
\label{f6}
\end{figure}
Note that after performing the required differentiations
one must set $\xi=1$. In Fig.~\ref{f6} (upper panel)
the obtained DFs are shown in the isotropic (solid lines) and strongly
anisotropic (dotted lines) cases. Again, inconsistency due to OM anisotropy
affects the DF at intermediate energies. The true anisotropy limits
derived from the DF are given by solid squares in Fig.~\ref{f2}.
Finally, note also the characteristic log-log linear trend of the DF
for very high values of $Q$. This trend is easily explained by
considering the asymptotic expansion given in eq.~(\ref{asymBH}).

\subsection{The halo dominated ($n$-1,1) Model}

We now derive the explicit DF for the family of halo dominated
($n$-1,1) models, so that with this class of models we can study the
effect of the external logarithmic slope when the two density profiles
have the same inner slope. Of course, the (4-1,1) models are the halo
dominated of (1,1) models discussed in C96.

Integration of eq.~(\ref{eq:fnu}) reveals that the functions involved
in the expression of the DF are the same as for the ($n$-2,1) models.
In particular,
\begin{eqnarray}\label{fisg1}
\Ftilis(\nu)&=&\displaystyle\frac{(-1)^{n}\mu\sqrt{1+\nu}\,
\xi^{n-3}}{2\pi(n-4)!}\,\nonumber\\
&&\displaystyle\frac{d^{n-4}}{d\xi^{n-4}}
\left(\frac{G^0_2-G_2}{\xi^{3}}-\frac{2G_3}{\xi^2}-\frac{3G_4}{\xi}\right),
\end{eqnarray}
and
\begin{equation}\label{fang1}
\Ftilan(\nu)=\displaystyle\frac{(-1)^{n}\mu\sqrt{1+\nu}\,\xi^{n-3}}{2\pi(n-4)!}\,
\displaystyle\frac{d^{n-4}(2 G_3-3\xi G_4)}{d\xi^{n-4}}.
\end{equation}
In Fig.~\ref{f5} (middle panel), the DF in the isotropic and strongly
anisotropic cases is shown, for $n=4,5,6$. The behaviour is
qualitatively similar to that of ($n$-2,1) models, i.e. while the isotropic
systems display a monotonically decreasing DF, in the anisotropic case
(when $\sacm$ approaches the consistency limit) the DF has a significant
depression, that moves inwards for increasing $n$.  In Fig.~\ref{f3}
(middle row) we show the minimum value of the anisotropy radius
normalized to $\rc$ as derived from the DF (dashed lines), and from
the WSC. Again, in accordance with the preliminary study of Section~3.2, 
we found that $\fis>0$ independently of the relative concentration value $\xi$ 
and of the external slope $n$, so that only the limit $\sacm$ exists.  
All the trends exhibited by the ($n$-2,1) models are confirmed: 
in particular, $\sacm$ increases with $\xi$ for
fixed $n$, while it decreases at fixed $\xi$ for increasing $n$.  Note
that the maximum radial anisotropy admissible for given $\xi$ and $n$ is
smaller than for ($n$-2,1) models, as expected from the shallower
central profile of $n$-1 models.  Finally, at variance with the
($n$-2,1) models (and in agreement with the expectations of the
preliminary analysis) no purely radial orbital configurations can be
supported, independently of the value of the concentration $\xi$.

As for ($n$-2,BH) models, also in the case of the black hole dominated
($n$-1,BH) models the functions $\Ftilis$ and $\Ftilan$ are formally
identical to those in eqs.~(\ref{fisg1})-(\ref{fang1}), with each $G$
function replaced by the corresponding $G_{\bullet}$ function, and
with the substitution $\sqrt{1+\nu}\to\sqrt{\nu}$; again, it is
necessary to set $\xi=1$ after performing the required
differentiations. While the true limits on the anisotropy radius are
plotted in Fig.~\ref{f2} with the solid squares, the obtained DFs are
shown in the isotropic (solid) and strongly anisotropic (dotted) cases
in the bottom panel of Fig.~\ref{f6}; not unexpectedly, the
inconsistency due to anisotropy manifests itself at intermediate
energies.  Finally, the behaviour of the DF for very high values of
$Q$ is in accordance with the asymptotic expansion of the DF in
eq.~(\ref{asymBH}).

\subsection{The halo dominated ($n$-0,1) Model}

This case is expected to be the more complicated, because from C99 it
is known that the (0,1) model presents a peculiar behaviour: in
practice, there exists a range of $\xi$ where anisotropic models can
be consistent when instead isotropic models are not, and this is due
to the fact that $\fis$ becomes negative over the non empty set $A_-$,
so that two limits on $\sa$ must be considered.  Note that the
discussion of the WSC in the case of ($n$-0,1) models in Section 3.2
already showed a similar behaviour.

As in the other cases, the integration of eq.~(\ref{eq:fnu})
can be done by using the $G_k$ functions only:
\begin{eqnarray}\label{fIg0}
\Ftilis(\nu)&=&\displaystyle\frac{(-1)^{n}3\mu\sqrt{1+\nu}\,\xi^{n-3}}
{\pi(n-4)!}\,
\frac{d^{n-4}G_5}{d\xi^{n-4}}\nonumber\\
&=&{n\choose4}\displaystyle\frac{3\mu\sqrt{1+\nu}\,\xi^{n-3}G_{n+1}}{\pi}\, ,
\end{eqnarray}
\begin{eqnarray}
\Ftilan(\nu)&=&\displaystyle\frac{(-1)^{n}3\mu\sqrt{1+\nu}\,\xi^{n-3}}
{2\pi(n-4)!}\,\nonumber\\
&&\displaystyle\frac{d^{n-4}(2\xi^2 G_5-3\xi G_4+G_3)}{d\xi^{n-4}},
\end{eqnarray}
where the second expression for $\Ftilis$ is obtained from the
relation (\ref{recG}) of the $G_{k}$ functions.  Motivated
by the remarks above, we start to study the sign of $\fis$, in order
to determine the condition for the existence of the sets $A_+$ and
$A_-$.  Indeed, we found that for $\xi\lsim 5.233$ (in the $n=4$
case, see C99), $\xi\lsim 6.192$ (for $n=5$), and $\xi\lsim 7.166$
(for $n=6$), $\fis$ is positive everywhere over the accessible
phase--space, a situation which is graphically represented with the
dotted lines in Fig.~\ref{f4}. It follows that for sufficiently
concentrated models only the anisotropy limit $\sacm$ exists, 
and in Fig.~\ref{f5} (bottom panel, solid lines) 
the isotropic DF is shown for ($n$-0,1) models
with $\xi=5$ and $n=(4,5,6)$: note how, at variance with the other
cases, the DF of ($n$-0,1) models decreases for increasing $Q$. 
The trend of $\sacm(\xi)$ is shown in Fig.~\ref{f3} (top panels, dashed
lines): as for the models previously discussed, the $\sacm$ curve is similar (but
displaced) with respect to that derived from the WSC. It is apparent 
how the $\sacm$ line lies above those of $n$-1 and $n$-2 models, as expected
from the shallower density profile of $n$-0 models; moreover, we note how the
minimum admissible values for $\sacm$ decrease for increasing $n$, as
in the other cases.  In Fig.~\ref{f5} the DFs for $\sa$ near the
consistency limit $\sacm$ are shown with dotted lines for models 
having $\xi=5$, and as in the previous $\gamma=1$ and $\gamma=2$ models,
inconsistency appears at intermediate values of $\Qtil$.

For values of $\xi$ larger than the critical values of $\xi_{\rm{c}}$
reported above, the set $A_-$ is not empty, because the isotropic
component of the DF becomes negative at high relative energies.  In
fact, at variance with the ($n$-2,1) and ($n$-1,1) models, the DF of
the $n$-0 model converges to a finite limit (that can be easily
calculated analytically from eq.~[\ref{fIg0}]) for $\Qtil=1$; 
when increasing $\xi=\rh/\rs$, the value $\ftil(1)$ monotonically decreases, 
till it becomes negative for $\xi$ greater than $\xi_{\rm{c}}$. 
A similar behavior was found in the numerical investigation
of consistency of King (1972) and quasi--isothermal halos added to a
de Vaucouleurs (1948) density distribution (see CP92).
Thus, while the effects of OM radial anisotropy appear to affect the
DF at intermediate energies, the inconsistency of isotropic models
embedded in a peaked halo seems to characterize the DF at very high
(relative) energies. According to the discussion of Section~2,
since the function $\fan$ is positive on $A_-$,
the upper bound $\sacp$ can be determined: 
the appearence of this new value is manifested by the
dotted lines in Fig.~\ref{f3}, with the vertical asymptote
corresponding to $\xi_{\rm{c}}$. Then, for $\xi$ greater than the
critical limit, a range of concentrations exists where an anisotropy
radius $\racm\leq\ra\leq\racp$ corresponds to consistent models;
finally, for $\xi$ larger than the value where the dotted ($\sacp$)
and dashed ($\sacm$) lines cross each other, $\sacm>\sacp$, and the
models become inconsistent.  Summarizing, consistent models correspond
to points placed above the $\sacm$ line and on the left of the $\sacp$ line.

In the case of a central dominant black hole, the same comments as
those for ($n$-2,BH) and ($n$-1,BH) models apply, and the functions
$\Ftilis$ and $\Ftilan$ are obtained accordingly.  Of course, the
isotropic model is inconsistent, independently of the values of $n$,
because the model with black hole would correspond to the formal limit
$\xi\to\infty$ in ($n$-0,1) models, which is already excluded by the
previous discussion.  What happens when considering the effect of
anisotropy?  Is there any compensating effect?  A numerical inspection
shows that the anisotropic part of the DF is negative on $A_-$, so
that not even orbital anisotropy can make the models consistent.
Again, this result was expected from the study of ($n$-0,1) models,
due to the crossing of the $\sacm$ and $\sacp$ critical lines at
finite values of $\xi$.

\section{Discussion and conclusions}

In this paper, in a natural extension of previous investigations
(CP92, C96, C99), we focused on the importance of the
\textit{external} density logarithmic slope on the determination of
the phase--space consistency of one and two--component stellar systems
with OM orbital anisotropy.  In particular, the considered
($\ngam$,$\gau$) models generalize ($\gau$,$\gad$) models, as the
\ngamma density component is characterized by logarithmic slope $n$
instead of $4$ outside the core radius, while in the inner regions the
logarithmic density slope is $\gamma$.  The main results can be
summarized as follows:

\begin{enumerate}

\item It is shown that, for $n$ integer, several structural and
  dynamical properties of the \ngamma models can be obtained by
  differentiation, with respect to the scale--length, of the
  corresponding formulae valid for the standard $\gamma$ models.  For
  example, it is shown how to construct explicit solutions of the
  Jeans equations for the class of two--component anisotropic
  ($n_1$-$\gau$,$n_2$-$\gad$) models once the solution is known for
  ($\gau$,$\gad$) models.

\item In one-component \ngamma models, a lower limit for the
  anisotropy radius, so that smaller values certainly produce
  inconsistent models, is analytically derived following the technique
  introduced in CP92. It is found that for $0\leq\gamma<2$ 
  this minimum anisotropy radius is strictly positive, but it decreases by increasing $n$,
  i.e. a larger amount of radial anisotropy can be supported by
  externally steeper density profiles. For $2\leq\gamma<3$ instead
  the necessary condition for consistency is satisfied $\forall\sa>0$,
  independently of the value of $\gamma$ and $n$.  The minimum
  anisotropy radius so that larger values correspond to consistent 
  \ngamma models is then determined by using the strong and weak sufficient conditions
  for model consistency derived in CP92. As for the necessary condition,
  we find that the minimum value of the anisotropy radius decreases for increasing
  $\gamma$ and fixed $n$, and for increasing $n$ and fixed $\gamma$.

\item A similar analysis is then performed for two-component
  ($\ngam$,$\gau$) systems, in order to extend the study of C99.  For
  simplicity we restrict to halo dominated models, i.e. we take into account 
  only the gravitational field of the $\gau$ halo.
  In particular, we show that in the isotropic case, when
  $1\leq\gamma<3$ and $0\leq\gau\leq\gamma$ (i.e., the halo density is
  centrally less peaked than the stellar component), the DF of the
  \ngamma component is nowhere negative, independently of the mass and
  concentration of the $\gau$ halo.  This is true even when in the
  external parts the \ngamma component is less peaked $(3<n<4)$ than
  the $\gau$ component. As a special application of this result, it
  follows that a black hole of any mass can be consistently added at
  the center of a globally isotropic \ngamma model, when $1\leq\gamma
  <3$. Moreover, in the case of ($n$-0,1) models, where instead the halo
  is centrally steeper than the stellar component, the sufficient
  condition applied to the $n$-0 density distribution reveals the
  existence of an upper limit $\xi\leq(n+1)/2$ of the relative
  concentration $\xi=\rh/\rs$ for model consistency in the isotropic
  case.  Finally, in the case of anisotropic \ngamma models with a
  dominant black hole at their center, we determined analytically a
  lower limit of the critical anisotropy radius for consistency as a
  function of $n$ and $\gamma$; this value decreases for increasing
  $\gamma$ and/or $n$.

\item The analytic expression for the DF of OM anisotropic
  halo dominated ($\ngam$,1) models, and (\ngam,BH) models with a
    dominant central BH is recovered in terms of elementary functions
  in the special cases when $\gamma=(0,1,2)$.  It is found that, while for
  $\gamma=1,2$ the isotropic DF is positive independently of the model
  concentration, the $\gamma=0$ component becomes inconsistent when
  the halo is sufficiently concentrated, even in the isotropic case.
  In addition, the trend of the minimum value of the anisotropy radius
  as a function of the halo concentration confirms that steeper
  density profiles in the external region are consistent with a larger
  amount of radial anisotropy.
\end{enumerate}

In conclusion, the explored family of models provide a direct
indication that the density slope in the external regions of stellar
systems can be important in determining the admissible radial
anisotropy, and this in addition to the well known relevance of the
central density slope.  A quantification of this argument is embodied
in a {\it necessary} inequality for model consistency, that we derived
by expressing a previous finding (CP92) in terms of the logarithmic
density slope.  The obtained result holds for {\it all} consistent OM
anisotropic systems, and relates the logarithmic density slope and the
OM anisotropy indicator at {\it each radius}, in a way formally
identical to the ``cusp slope-central anisotropy'' theorem by AE06,
which is known to apply at the center of stellar systems with
generic anisotropy distribution, and everywhere in constant anisotropy
systems.

\section*{Acknowledgments} We wish to thank the anonymous Referee for 
helpful suggestions that greatly improved the presentation of the paper.

\appendix

\section{Consistency requirements}

\subsection{The NC for one--component \ngamma models}

Simple algebra shows that the NC applied to anisotropic \ngamma models
can be written as
\begin{equation}
\sa^2\geq \frac{2-\gamma-(n-2)s}{\gamma+ns}s^2, \quad  0\leq
s\leq\infty .
\end{equation}
For $\gamma\geq2$ and $n\geq2$ the r.h.s. is everywhere negative,
so that inequality (A1) is trivially satisfied for $\sa\geq0$.
Instead, for $0\leq\gamma<2$ the parameter $\sa$
must be larger than the maximum of the expression of the r.h.s.,
which is reached at
\begin{equation}
\sM (n,\gamma) =\frac{n-(2n-3)\gamma+
                 \sqrt{(n-\gamma)[n+(4n-9)\gamma]}}{2n(n-2)},
\end{equation}
that after substitution in eq.~(A1) gives eq.~(\ref{eq:ngNC}).
Of course, for $n=4$ eq.~(A2) in C99 is reobtained.

\subsection{The WSC for halo--dominated (\ngam,$\gau$) models}

Here we apply the WSC to the \ngamma component of globally isotropic,
halo--dominated (\ngam,$\gau$) models, where
$1\leq\gamma <3$, $0\leq\gau\leq\gamma$, and $n>3$.  Under the
assumption of a dominant halo, eq.~(\ref{eq:WSC}) reduces to
investigate the positivity of an expression which factorizes in a
strictly positive function and in an algebraic factor that after the
natural substitutions $\gamma=1+\epsilon$ and $\gau=\gamma-\epu$ (with
$0\leq\epsilon <2$ and $0\leq\epu\leq\gamma$) becomes
\begin{eqnarray}
&&n(n-1)s^3+n[n+1-\epsilon+\epu+2\xi\epsilon]s^2\\
&&+\xi\{[(n+1)(2+\epsilon+\epu)
+\epsilon\epu-\epsilon^2]+\xi\epsilon\gamma\}s+
                      \xi^2\gamma(2+\epu),\nonumber
\end{eqnarray}
whose positivity $\forall (s,\xi)\geq 0$ is easily proved.  Therefore
the WSC is verified for any value of the concentration parameter
$\xi$.  Note that the variables $\xi$ and $s$ adopted here correspond
to $1/\xi$ and $s/\xi$ respectively in eq.~(A13) of C99, due to the
different normalization length.

Being the WSC of the isotropic models satisfied, the existence of a critical anisotropy
radius $\sacm$ for consistency depends on the sign of the anisotropic
part of the WSC, as discussed in Section~2.2. 
In particular, the positivity of the anisotropic WSC for ($n$-2,1) models reads
\begin{equation}
s(n-3)+n-1-2\xi\geq 0,\quad 0\leq s\leq\infty,
\end{equation}
and thus shows the existence of a critical value $\xi_{\rm{c}}=(n-1)/2$
of the concentration parameter, marking an upper limit on the values
of $\xi$ for which $\sacm=0$, i.e. purely radial orbital distributions
are allowed.

The discussion of the WSC for globally isotropic $n$-0
models in a dominant $\gamma=1$ halo is simple,
reducing to the request that
\begin{equation}
s(n-1)+n+1-2\xi\geq 0,\quad 0\leq s\leq\infty,
\end{equation}
which is satisfied for $\xi\leq\xi_{\rm{c}}=(n+1)/2$.
For $n=4$ we reobtain the result of C99.

In the case of central dominant BH, we assume in eq.~(\ref{eq:WSC})
$\mt=M_{\rm BH}$, and from the previous discussion we restrict
to \ngamma models with $1\leq\gamma <3$.  
The WSC then requires that
\begin{equation}
\sa^2\geq
s^2\frac{(3-\gamma)(\gamma-2)+2(n-2)(3-\gamma)s-(n-2)(n-3)s^2}{
              n(n-1)s^2+2n(\gamma-1)s+\gamma(\gamma-1)}.
\end{equation}
After the differentiation of the r.h.s. and the successive study 
of a quartic equation, it can be proved that for $s\geq0$
eq.~(A6) admits only one maximum, located at $\sM=\sM(n,\gamma)\geq 0$.
The general expression for $\sM(n,\gamma)$ is not reported here,
but it was used to produce the long--dashed lines in Fig.~\ref{f2}.
In any case, $\sM(n,3)=0$, 
\begin{equation}
\sM(n,\gamma)=
\begin{cases}
\displaystyle{\frac{2}{n-3}},
               &\gamma=1,\cr
               \displaystyle{\frac{\sqrt[3]{s_0}}{3n(n-1)(n-3)}-
         \frac{2(n-4)}{3(n-1)(n-3)}}\cr
         +\displaystyle{\frac{4(n^3+n^2-20n+27)}{3 (n-1)(n-3)\sqrt[3]{s_0}}},
               &\gamma=2,\cr
\end{cases}
\end{equation}
where
\begin{eqnarray}
s_0&=&n^2\left[73n^4-660n^3+2262n^2-3484n+2025\right.\nonumber\\
&+&\left.9\sqrt{\frac{(5n-9)(13n^2-59n+64)(n-1)^3(n-3)^3}{n}}\right],
\end{eqnarray}
and for $n=4$ and $\gamma=1,2,3$ the values given in Table 1 of C99
are reobtained.
We finally report the reduced asymptotic fourth--order
equation needed to determine $\sM(n,\gamma)$ in the limit $n\to\infty$:
remarkably, after the scaling $s=y/n$ and the successive limit $n\to\infty$,
one obtains
\begin{eqnarray}
&y^4&+2(2\gamma-3)y^3+6(\gamma-1)(\gamma-2)y^2\\
&+&2(\gamma-3)(\gamma-1)(2\gamma-1)y+(\gamma-3)(\gamma-2)(\gamma-1)\gamma=0.\nonumber
\end{eqnarray}

\section{The velocity dispersions and the virial quantities for
  \ngaugad models}

Here we present an easy way to express analytically the main dynamical
quantities of the two--component OM anisotropic
($n_1$-$\gau$,$n_2$-$\gad$) models for generic $\gau$ and $\gad$, and
$n_1$ and $n_2$ integer $\geq 4$.  The method is based on the
evaluation of the similar quantities for the two--components
$(\gau,\gad)$ models, and then on repeated differentiation with
respect to the scale--lengths.

We start by considering the radial component $\sigr^2$ of the velocity
dispersion $\sigma^2=\sigr^2+\sigt^2$ of a system with a density
component $\rho$, which in the OM parameterization can be written as
\begin{equation}\label{bin&mer}
\rho(r)\sigr^2(r)=\frac{A(r)+\ra^2 I(r)}{r^2+\ra^2},
\end{equation}
where
\begin{equation}
A(r)=G\int_r^{\infty}\rho(r)\mt(r)dr,\quad
I(r)=G\int_r^{\infty}\frac{\rho(r)\mt(r)}{r^2}dr,
\end{equation}
(Binney \& Mamon 1982), and $\mt(r)$ is the total mass within $r$.
Once $\sigr^2$ is known, the tangential velocity
dispersion is obtained from eq.~(\ref{betaOM}).

As the method is general, we focus without loss of generality on the
component ``1'' of a two--component model, for which the two functions
$I$ and $A$ are given by the sum $\Is =\Iss+\Ish$ and $\As=\Ass+\Ash$,
due to the linearity of mass in eqs.~(B2), and where the meaning of
the subscript indices is apparent.  Other quantities of interest in
applications are the global energies entering the scalar virial
theorem, and for the component 1 the scalar virial theorem reads
$2K_1=|W_{11}|+|\wsh|$, where $W_{11}=-4\pi G\int r\rhos\ms dr$ is the
contribution due to the self--interaction, and $\wsh=-4\pi G\int
r\rhos M_{\rm2}dr$ is the interaction energy with the ``halo'' (e.g.,
Ciotti 2000).

The basic idea is to compute the integrals for the generic pair
($\rhos,M_{\rm2}$) relative to standard two--component $(\gau,\gad)$
models with different scale--lengths $\rcu$ and $\rcd$, to perform the
required differentiations as prescribed by eqs.~(\ref{eq:ng}) and (\ref{eq:Mng}), 
and finally (in the case of self--interactions) to set the two
scale--lengths and the two slopes to the same value. 
From eqs.~(\ref{eq:ng}) and (\ref{eq:Mng}) it follows
\begin{eqnarray}
A_{12}(r)&=&G\frac{\rcu^{n_1-3}\rcd^{n_2-3}(-1)^{n_1+n_2}}
{\Gamma(n_1-3)\Gamma(n_2-3)}\nonumber\\
               &&\frac{d^{n_1-4}}{d\rcu^{n_1-4}}\frac{d^{n_2-4}}{d\rcd^{n_2-4}}
               \int_r^{\infty}\frac{\rho_{\gau}(r,\rcu)M_{\gamma_2}(r,\rcd)}
{\rcu\rcd}dr,\\
I_{12}(r)&=&G\frac{\rcu^{n_1-3}\rcd^{n_2-3}(-1)^{n_1+n_2}}
{\Gamma(n_1-3)\Gamma(n_2-3)}\nonumber\\
               &&\frac{d^{n_1-4}}{d\rcu^{n_1-4}}\frac{d^{n_2-4}}{d\rcd^{n_2-4}}
               \int_r^{\infty}\frac{\rho_{\gau}(r,\rcu)M_{\gamma_2}(r,\rcd)}
{\rcu\rcd r^2}dr,\\
W_{12}(r)&=&-4\pi G\frac{\rcu^{n_1-3}\rcd^{n_2-3}(-1)^{n_1+n_2}}
{\Gamma(n_1-3)\Gamma(n_2-3)}\nonumber\\
               &&\frac{d^{n_1-4}}{d\rcu^{n_1-4}}\frac{d^{n_2-4}}{d\rcd^{n_2-4}}
               \int_0^{\infty}\frac{\rho_{\gau}(r,\rcu)M_{\gamma_2}(r,\rcd)}
{\rcu\rcd}r dr;
\end{eqnarray}
for generic values of $\gau$ and $\gad$ the integrals above involve
hypergeometric ${}_2F_1$ functions, while for integer values of $\gau$
and $\gad$ the resulting formulae can be expressed in terms of
elementary functions (e.g., see C96 and C99 for (1,0) and (1,1)
models; Ciotti et al. 1996 for (2,1) and (2,2) models).  Even
  simpler expressions can be obtained without difficulty for to the
  halo dominated ($n$-0,1), ($n$-1,1), and ($n$-2,1) models, 
by adopting as normalization constants the physical
  scales $\ms$ and $\rs$ of the $\gamma=1$ dominant component, and by
  fixing $n_2=4$, $\rcd=\rs$, $n_1=n$, $\xi\equiv\rcu/\rs$, $s\equiv
  r/\rs$, $\mu\equiv M/\ms$ in eqs.~(B3)-(B6), where $M$ and $\rcu$
  are the mass and the scale--lenght of the $\ngam$ component. As the
  integration is trivial but the results not particularly
  illuminating, we do not show the explicit formulae.

\end{document}